\documentclass[draftcls,10pt,onecolumn]{IEEEtran}
\IEEEoverridecommandlockouts
\usepackage{graphicx}
\usepackage{amsmath}
\usepackage{amssymb}               
\usepackage{amsfonts}               
\usepackage{amsopn}

\allowdisplaybreaks
\begin{document}
\newtheorem{theorem}{\bf Theorem}
\newtheorem{lemma}{\bf Lemma}
\newtheorem{claim}{Claim}
\newtheorem{proposition}{\bf Proposition}
\newtheorem{conjecture}{Conjecture}
\newtheorem{corollary}{\bf Corollary}
\newtheorem{definition}{\bf Definition}
\newtheorem{example}{Example}
\newtheorem{discussion}{Discussion}
\newtheorem{remark}{\bf Remark}
\def\p{\mathsf{P}}
\def\hY{\hat{Y}}
\def\hy{\hat{y}}
\def\hz{\hat{z}}
\def\hu{\hat{u}}
\def\hZ{\hat{Z}}
\def\hS{\hat{S}}
\def\hs{\hat{s}}
\def\hR{\hat{R}}
\def\hp{\hat{P}}
\def\tp{\tilde{P}}
\def\tX{\tilde{X}}
\def\tY{\tilde{Y}}
\def\tZ{\tilde{Z}}
\def\tR{\tilde{R}}
\def\styp{\ms_{\gamma}(p\|t)}
\def\typ{\mathcal{T}_{\epsilon}^n}
\def\Reals{\mathbb R}
\def\n{\nonumber\\}
\def\be{\begin{equation}}
\def\ee{\end{equation}}
\def\bes{\begin{equation*}}
\def\ees{\end{equation*}}
\def\beq{\begin{eqnarray}}
\def\eeq{\end{eqnarray}}
\def\beqs{\begin{eqnarray*}}
\def\eeqs{\end{eqnarray*}}
\def\ma{{\mathcal A}}
\def\mb{{\mathcal B}}
\def\mc{{\mathcal C}}
\def\md{{\mathcal D}}
\def\me{{\mathcal E}}
\def\mf{{\mathcal F}}
\def\mg{{\mathcal G}}
\def\mh{{\mathcal H}}
\def\mi{{\mathcal I}}
\def\mj{{\mathcal J}}
\def\mk{{\mathcal K}}
\def\ml{{\mathcal L}}
\def\mm{{\mathcal M}}
\def\mn{{\mathcal N}}
\def\mo{{\mathcal O}}
\def\mp{{\mathcal P}}
\def\mq{{\mathcal Q}}
\def\mr{{\mathcal R}}
\def\ms{{\mathcal S}}
\def\mt{{\mathcal T}}
\def\mv{{\mathcal V}}
\def\mw{{\mathcal W}}
\def\mx{{\mathcal X}}
\def\my{{\mathcal Y}}
\def\mz{{\mathcal Z}}
\def\ua{\mathbf{a}}
\def\ub{\mathbf{b}}
\def\uc{\mathbf{c}}
\def\ud{\mathbf{d}}
\def\ue{\mathbf{e}}
\def\uf{\mathbf{f}}
\def\ug{\mathbf{g}}
\def\uh{\mathbf{h}}
\def\ui{\mathbf{i}}
\def\uj{\mathbf{j}}
\def\uk{\mathbf{k}}
\def\ul{\mathbf{l}}
\def\um{\mathbf{m}}
\def\un{\mathbf{n}}
\def\uo{\mathbf{o}}
\def\br{\mathbf{R}}
\def\bC{\mathbf{C}}
\def\F{\mathbf{F}}
\def\bp{\mathbf{P}}
\def\uq{\mathbf{q}}
\def\ur{\mathbf{r}}
\def\us{\mathbf{s}}
\def\ut{\mathbf{t}}
\def\uu{\mathbf{u}}
\def\uv{\mathbf{v}}
\def\uw{\mathbf{w}}
\def\e{\mathbb{E}}
\def\tt{\mathbf{1}((u^n,v^n,w^n,y^n,z^n)\in\styp)}
\def\tv#1{\left\|#1\right\|_1}
\def\apx#1{\stackrel{#1}{\approx}}
\def\ux{x^n}
\def\uy{y^n}
\def\uz{z^n}
\def\uyy{\mathbf{y}}

\def\uhy{\mathbf{\hat{y}}}
\def\vanish{\xrightarrow{\footnotesize exp} 0}
\def\mathllap{\mathpalette\mathllapinternal}
 \def\mathllapinternal#1#2{%
 \llap{$\mathsurround=0pt#1{#2}$}
 }
 \def\clap#1{\hbox to 0pt{\hss#1\hss}}
 \def\mathclap{\mathpalette\mathclapinternal}
 \def\mathclapinternal#1#2{%
 \clap{$\mathsurround=0pt#1{#2}$}%
 }
 \def\mathrlap{\mathpalette\mathrlapinternal}
 \def\mathrlapinternal#1#2{%
 \rlap{$\mathsurround=0pt#1{#2}$}
 }
\def\Sum{\ensuremath\mathop{\scalebox{1.2}{$\displaystyle\sum$}}}
\def\Prod{\ensuremath\mathop{\scalebox{1.2}{$\displaystyle\prod$}}}
\def\im{\imath}
\def\s#1{\mathsf{#1}}
\def\ep{\epsilon}
\def\Var{\mathsf{Var}}
\def\TV{\mathsf{TV}}
\def\sec{\mathsf{sec}}
\def\Cov{\mathsf{Cov}}
\def\SW{\mathsf{SW}}
\def\ap{\mathsf{Apx}}
\def\sm{\mathsf{M}}
\def\sf{\mathsf{F}}
\def\sj{\mathsf{J}}
\def\enc{\mathsf{Enc}}
\def\dec{\mathsf{Dec}}
\sloppy

\author{Mohammad~Hossein~Yassaee, Mohammad~Reza~Aref, Amin~Gohari\\Information Systems and Security Lab (ISSL),\\Sharif University of Technology, Tehran, Iran,\\
E-mail: yassaee@ee.sharif.edu, \{aref,aminzadeh\}@sharif.edu.
 \thanks{ This work was supported by Iran-NSF under grant No. 88114.46.}}

\title{\Huge Non-Asymptotic Output Statistics of Random Binning and Its Applications}

\maketitle

\begin{abstract}
In this paper we develop a finite blocklength version of the Output Statistics of Random Binning (OSRB) framework. The framework is shown to be optimal in the point-to-point case. 
New second order regions for broadcast channel and  wiretap channel with strong secrecy criterion are derived.
\end{abstract}
\section{Introduction}
Output Statistics of Random Binning (OSRB) is a new framework for proving achievability results \cite{OSRB}. It works by converting channel coding problems into source coding problems, and uses the induced pmf of the source coding side to design encoders for the channel coding side. The goal is to make the total variation distance between the joint pmf of the source coding side and channel coding side close to zero so that all the performance analysis can be dealt with at the source coding side where Slepian-Wolf (S-W) theorem can be invoked. Thus the OSRB technique is not based on the usual covering and packing lemmas.

Originally studied by Strassen \cite{strassen}, there has been a recent surge of works on \emph{finite blocklength information theory} following the work of Polyanskiy et al \cite{PPV} (see for instance \cite{verdual}-\cite{macf}). 
In this paper we develop a finite blocklength version of the OSRB framework. We show that this method is optimal in the point-to-point channel and can directly give us the channel dispersion $\e[\Var(\im(X;Y))|X]$.\footnote{Direct proofs for this formula have also been obtained by Wang et. al. using a different technique \cite{kochman}.} We also use the technique to derive the second order region for broadcast channel (that recovers Marton's inner bound in the asymptotic case) and for wiretap channel with strong secrecy criterion (that improves the  result of \cite{tanwt}). Scenarios such as broadcast wiretap channel can be also dealt with using this technique but have been left out for a more complete version of this draft.

OSRB is based on two theorems: the S-W theorem and another theorem that may be considered as its dual. To develop a finite blocklength version of the OSRB, we first find a one-shot version of these two main theorems. By one-shot we mean that only a single use of the resource is allowed. To get finite blocklength results, we then apply this result to a product of $n$ use of the network. The resulting dispersion at the output can be either due to the dispersion in the input code or to the inherent dispersion of the channel. To avoid the input dispersion, we use a uniform distribution over a fixed type in the source coding side of the problem. In this sense this differs from the original asymptotic OSRB where we use a completely i.i.d. distribution in the source coding side of the problem. 

This paper is organized as follows: some definitions and notations are given in Section \ref{sec:1}. One-shot version of the two main theorems of the OSRB are given in Section \ref{sec:2}. We then apply the technique to a couple of problems in Section \ref{sec:3}. To illustrate the use of the technique we begin by recovering the known result on dispersion for the point-to-point channel in Subsection \ref{sec:31}. In Subsections \ref{sec:32} and \ref{sec:33}, we apply the technique to broadcast channel and wiretap channel.
\section{Definitions}\label{sec:1}\begin{definition}
Given a pmf $p_{X,Y}$, the conditional information of $x$ given $y$ is defined by
\[h_p(x|y):=\log\frac{1}{p_{X|Y}(x|y)}.\]
Also, the \emph{information density} $\im_p(x;y)$ is defined by
\[ \im_p(x;y):=\log \frac{p(x,y)}{p(x)p(y)}.\]
\end{definition}
\begin{definition}
Let $\mathbf{X}$ be a multi-dimensional normal variable with zero mean and covariance matrix $\mathsf{V}$. The complementary multivariate Gaussian cumulative distribution region associated with $\mathsf{V}$ is defined by
 \[
\mq^{-1}(\s{V},\epsilon):=\{\mathbf{x}:\s{P}(\mathbf{X\le x})\ge 1-\ep\}.
\]
\end{definition}
\par \textbf{Notation}: In this paper, 
we use $X_{\mv}$ to denote $(X_v:v\in\mv)$ and
$p^U_{\ma}$ to denote the uniform distribution over the set $\ma$. The total variation between two pmf's $p$ and $q$ on the same alphabet $\mx$ , is defined by $\tv{p(x)-q(x)}:=\frac{1}{2}\sum_x|p(x)-q(x)|$. When a pmf itself is random, we use capital letter, e.g. $P_X$. See \cite[Remark 1]{OSRB} for more details about random pmfs and their manipulations. \section{One-shot Output Statistics of Random Binning}\label{sec:2}
Let $(X_{\mv},Z)$ be a set of discrete sources distributed according to a joint pmf $p_{X_{\mv},Z}$ on a finite set $(\prod_{v\in \mathcal{V}} \mx_v)\times \mz$. A distributed  random  binning consists of a set of random mappings $\mb_v: \mx_v\rightarrow [1:\sm_v]$, $v\in\mv$, in which $\mb_v$ maps each sequence of $\mx_v$ uniformly and independently to the set $[1:\sm_v]$. We use  $B_v$ as a shorthand for rv $\mb_v(X_v)$, and $B_{\mv}$ or $B_{\mv}(X_{\mv})$ as a shorthand for rv $(\mb_v(X_v))_{v\in\mv}$. A  distributed random binning induces the following \emph{random pmf}\footnote{The pmf is random due to the random binning assignment in the protocol.} on the set $\mx_{\mv}^n\times\mz^n\times\prod_{v\in \mathcal{V}} [1:\sm_v]$,
\begin{align}\label{eq:OSRBS1}
P(x_{\mv},z,b_{\mv})=p(x_{\mv},z)\prod_{v\in \mathcal{V}}\mathbf{1}\{\mb_v(x_v)=b_v\}.
\end{align}
The asymptotic OSRB introduced in \cite{OSRB} relies on the S-W theorem as well as Theorem 1 of \cite{OSRB} that implies independence of random bin indices under certain conditions. To set up a \emph{non-asymptotic} framework, we generalize the S-W theorem and Theorem 1 of \cite{OSRB} to the case of a single channel use. Let us begin with the latter:
\begin{theorem}\label{thm:os1}
Given $p_{X_{\mv},Z}$, for any  pmf $t_{Z}$ and any positive real $\gamma$,  the random pmf of eq. \eqref{eq:OSRBS1} satisfies
\begin{align*}
\e\tv{P(b_{\mv},z)-p^U(b_{\mv})p(z)}\le p_{X_{\mv}Z}(\ms_{\gamma}(p\|t)^c)+2^{\frac{|\mv|-\gamma}{2}-1},
\end{align*}
where the expectation is over the randomness of binning and the set $\ms_{\gamma}(p\|t)\subset \mx_{\mv}^n\times\mz^n$ is defined as follows:
\begin{align*}
\ms_{\gamma}(p\|t):=\{(x_{\mv},z):\forall \emptyset\neq\ms\subseteq&\mv, h_p(x_{\ms},z)-h_t(z)-\sum_{v\in\ms}\log \sm_v>\gamma\}.
\end{align*}
\end{theorem}
\begin{remark} This theorem implies  \cite[Theorem 1]{OSRB} by evaluating it for the product $p_{X_{\mv}^n,Z^n}=\prod_{i=1}^np_{X_{\mv,i},Z_i}$.  Set $t_{Z^n}=p_{Z^n}$ and $\gamma=n\delta$ for a sufficiently small value of $\delta>0$ that we discuss later. Then the term $2^{\frac{|\mv|-1-\gamma}{2}}$ converges to zero. The first term converges to zero as well if $\ms_{\gamma}(p\|t)$ includes almost all of the typical set  $\mt_{p}$. For any jointly typical $(x^n_{\mv},z^n)$, the terms $h_p(x_{\ms}^n,z^n)$ and $h_p(z^n)$ are close to $nH_p(X_{\ms},Z)$ and $nH_p(Z)$, respectively. Thus as long as $H_p(X_{\ms}|Z)>\frac{1}{n}\sum_{v\in\ms}\log M_v$ we can choose $\gamma=n\delta$ for a $\delta>0$ such that the inequalities defining  $\ms_{\gamma}(p\|t)$ holds.
\end{remark}
\begin{remark}The rv $Z$ in the statement of the above theorem is of use in problems with secrecy constraints. 
\end{remark}
\begin{IEEEproof}
See Appendix \ref{apx:nosrb}.
\end{IEEEproof}

\emph{One shot S-W coding:}
Here we want to bound the error probability of decoding a single copy of the source $X_{\mv}$ when the decoder has access to the side information $Z$ as well as the bin indices $B_{\mv}$. An optimal decoder uses ML decoding. However we use an \emph{stochastic} variation of MAP for the decoding with a more tracktable analysis. The decoder draws $\hat{x}_{\mv}$ from the conditional pmf $P_{X_{\mv}|Z,B_{\mv}}(\hat{x}_{\mv}|y,b_{\mv})$, where $P$ is the induced probability by the random binning. More specifically
\bes
P_{X_{\mv}|Z,B_{\mv}}(\hat{x}_{\mv}|z,b_{\mv})=\dfrac{p(\hat{x}_{\mv}|z)\mathbf{1}(B_{\mv}(\hat{x}_{\mv})=b_{\mv})}{\sum_{\bar{x}_{\mv}}p(\bar{x}_{\mv}|z)\mathbf{1}(B_{\mv}(\bar{x}_{\mv})=b_{\mv})}.
\ees 
We refer this decoder as a \emph{stochastic likelihood coder} (SLC). See \cite{13} for a motivation of SLC and the justification for using a stochastic decoder. For some technical reasons,we can  more generally use a \emph{mismatch} SLC corresponding to an arbitrary pmf $t_{X_{\mv},Z}$ instead of $p$ in the above expression,\footnote{The pmf $t_{X_{\mv},Z}$ should not be confused with the one used in Thm \ref{thm:os1} where it is only defined on $Z$.} that is,
\bes
T_{X_{\mv}|Z,B_{\mv}}(\hat{x}_{\mv}|z,b_{\mv})=\dfrac{t(\hat{x}_{\mv}|z)\mathbf{1}(B_{\mv}(\hat{x}_{\mv})=b_{\mv})}{\sum_{\bar{x}_{\mv}}t(\bar{x}_{\mv}|z)\mathbf{1}(B_{\mv}(\bar{x}_{\mv})=b_{\mv})}.
\ees  
Roughly speaking, the reason for introducing a mismatch SLC is that we will need to work with input codewords of the same type to reduce the total dispersion, rather than with codewords generated from an i.i.d. distribution. However we need independence to be able to use the Berry-Essen CLT  at a later stage. A mismatch SLC allows us to simultaneously employ an independent and a non-independent  distribution.

\begin{theorem}\label{thm:os2}
Given $p_{X_{\mv},Z}$ and any pmf $t_{X_{\mv},Z}$, the expected value of the probability of correct decoding of a mismatch SLC associated with $t$ is bounded from below by
\be\label{eq:sw}
\e\mathsf{P}[C]\ge\e_{p}\dfrac{1}{1+\sum_{\emptyset\neq\ms\subseteq\mv}\sm_{\ms}^{-1}2^{h_t(X_{\ms}|X_{\ms}^c,Z)}},
\ee
where $\sm_{\ms}=\prod_{v\in\ms}\sm_v$. Moreover, this bound can be weakened to give the following bound on the error probability of mismatch SLC,
\begin{align}
\e \mathsf{P}[\me]\le& p_{X_{\mv}Z}(\ms_{\gamma}(t_{X_{\mv},Z})^c)+(2^{|\mv|}-1)2^{-\gamma},
\end{align}
where $\gamma$ is an arbitrary positive number and 
\begin{align}
\ms_{\gamma}(t_{X_{\mv},Z}):=\{(x_{\mathcal{V}},z):\forall \emptyset\neq\ms\subseteq\mathcal{V}, \sum_{v\in\ms}\log \sm_v-h_t(x_{\ms}|z)>\gamma\}.
\end{align}
\end{theorem}
\begin{remark} Using this theorem one can derive finite blocklength analogs of the S-W theorem for i.i.d. or non-i.i.d. sources. Since we choose the codewords from a fixed type, we use this theorem in its non-i.i.d. form. I.i.d. forms of the S-W theorem have been previously obtained by \cite{macf}.

\end{remark}
\begin{IEEEproof}
We only prove the inequality \eqref{eq:sw} for the special case of $|\mv|=1$. For the complete proof, see Appendix \ref{apx:sw}. The probability of correct decoding can be written as,
\[\mathsf{P}[C]=\sum_{x,b,z}p(x,z)\mathbf{1}(B(x)=b)T_{X|Z,B}(x|z,b).\]
We have,
\begin{small}
\begin{align}
&\e\mathsf{P}[C]=\e\sum_{x,z,b}p(x,z)\mathbf{1}(B(x)=b)
                         \dfrac{t(x|z)}{\sum_{\bar{x}}t(\bar{x}|z)\mathbf{1}(B(\bar{x})=b)}\\
                         &=\sm\e\sum_{x,z}p(x,z)\mathbf{1}(B(x)=1)
                         \dfrac{t(x|z)}{\sum_{\bar{x}}t(\bar{x}|z)\mathbf{1}(B(\bar{x})=1)}\label{eq:sw1}\\
                         &=\sm\sum_{x,z}\e_{B(x)}\e_{\{B(\bar{x}), \bar{x}\neq x\}}p(x,z)
                         \dfrac{t(x|z)\mathbf{1}(B(x)=1)}{\sum_{\bar{x}}t(\bar{x}|z)\mathbf{1}(B(\bar{x})=1)}\label{eq:sw2}\\
                         &\ge\sm\sum_{x,z}\e_{B({x})}p(x,z)
                         \dfrac{t(x|z)\mathbf{1}(B(x)=1)}{\e_{\{B(\bar{x}), \bar{x}\neq x\}}\sum_{\bar{x}}t(\bar{x}|z)\mathbf{1}(B(\bar{x})=1)}\label{eq:sw3}\\
                         &=\sm\sum_{x,z}\e_{B(x)}p(x,z)
                         \dfrac{t(x|z)\mathbf{1}(B(x)=1)}{t(x|z)\mathbf{1}(B(x)=1)+\sm^{-1}(1-t(x|z))}\label{eq:sw4}\\ 
                         &\ge\sm\sum_{x,z}\e_{B(x)}p(x,z)
                         \dfrac{t(x|z)\mathbf{1}(B(x)=1)}{t(x|z)\mathbf{1}(B(x)=1)+\sm^{-1}}\label{eq:sw4}\\                                                  
                         &=\sum_{x,z}p(x,z)\dfrac{t(x|z)}{t(x|z)+\sm^{-1}}=\e_{p}\dfrac{1}{1+\sm^{-1}2^{h_t(x|z)}},  \label{eq:sw6}                
\end{align}
\end{small}
where \eqref{eq:sw1} is due to the symmetry, \eqref{eq:sw3} follows from the Jensen inequality for the convex function $f(x)=\frac{1}{x}$ on the $\mathbb{R}_+$ and \eqref{eq:sw4} follows from the fact that $B(\bar{x})$ and $B(x)$ are independent for any $\bar{x}\neq x$.
\end{IEEEproof}
\section{Applications of non-asymptotic OSRB}\label{sec:3}
To illustrate the use of the tools introdued in the previous section, we recover a finite blocklength result for the point to point channel coding, and  prove new results for broadcast channel and wiretap channel. Since the structure of the proofs are similar, we have tried to provide a detailed proof for the simplest case, i.e. the point-to-point channel and outline other proofs have less details. See \cite{FullVersion} for the full proofs.

\subsection{Point to point channel coding}\label{sec:31}
\label{sub:ch}
Consider a DMC channel $q_{Y|X}$. We will recover the result of \cite{PPV} that there is an $(n,\epsilon)$-code with rate
\be R(n,\epsilon)=I(X;Y)+\sqrt{\dfrac{V}{n}}Q^{-1}(\ep)-O\left(\dfrac{\log n}{n}\right),\label{eq:ch1}\ee
 for any arbitrary input pmf $q_X$ 
 where $V=\e\left[\Var_{q_{Y|X}}(\im(X;Y)|X)\right]$. Our framework is divided into two steps: in the first step we obtain a one-shot achievable rate following the OSRB technique. In the second step we use Theorem \ref{thm:os1} and Theorem \ref{thm:os2} for the $n$ uses of the channel, to approximate the achievable rate. 
\subsubsection*{Step 1: One-shot OSRB}
Just like the asymptotic OSRB, the first step is itself divided into three parts. In the first part we start from a source coding problem, use random binning and then find an upper bound on the error probability. In the second part, we use the joint pmfs of the source coding side of the problem to design a concrete encoder-decoder for the channel coding with one exception: the encoder-decoder is assisted with a common randomness that does not really exist in the model (to be removed in third part). We will find upper bounds on the total variation distance of  the joint induced pmf's between all r.v.'s in the two parts. The bounds on the error probability of S-W coding and the total variation distance of  the joint induced pmf's give a bound on the error probability of encoder-decoder of the part two. In the third part, we eliminate the common randomness given to the second protocol without disturbing the probability of error. This makes the designed encoder-decoder in the second part useful for code construction.   
\subsubsection*{Part 1: Source coding problem and random binning} We start from a different problem of source coding; we will use the pmf induced by this problem to construct our channel code in the next part. Let $(X,Y)$ be distributed according to $q(x,y)=q(x)q(y|x)$. We define two random mappings on $\mathcal{X}$ as follows: to each $x$, we assign two random bin indices $m\in[1:\sm]$ and $f\in[1:\sf]$, uniformly and independently. This induces a joint pmf on $M,F,X$ which we denote by $P_s(m,f,x)$. Suppose that the decoder  chooses a $t_{X,Y}$ and uses a mismatched decoder $T(\hat{x}|y,f)$ constructed using $t_{X,Y}$. Then the induced random pmf is $P_s(x,y,m,f,\hat{x})=q(x,y)P_s(m,f|x)T(\hat{x}|f,y)$. Invoking Theorem \ref{thm:os2} with rv $Z$ being a constant, one can derive an upper bound $\epsilon_{\mathsf{Dec}}$ on the expectation of error probability that only depends on $\sf$ (and not on $\sm$). This upper bound is provided later in equation \eqref{eq:14} for the finite blocklength coding. 
\subsubsection*{Part 2: Designing encoder-decoder assisted with a shared randomness}
Returning to the channel coding problem we assume that there is a shared randomness $F$ available at both the encoder and decoder, which is independent of the message and uniformly distributed over $[1:\sf]$. This shared randomness does not exist in the original setup and we will eliminate it later. The encoder uses the conditional pmf $P_s(x|m,f)$ of the source coding problem. The decoder uses the mismatched decoder $T(\hat{x}|y,f)$ to find $\hat{x}$ and thereby an estimate of the message $\hat{m}$. The induced random pmf is $P_c(x,y,m,f,\hat{x})=p^U(m,f)P_s(x|m,f)p(y|x)T(\hat{x}|f,y)$. We have 
\begin{align}
\tv{P_s(x,y,m,f,\hat{x})-P_c(x,y,m,f,\hat{x})}=\tv{P_s(m,f)-p^{U}(m,f)}.\label{eq:A1}\end{align}

Given $\sm$ and $\sf$, Theorem \ref{thm:os1} gives an upper bound $\epsilon_{\mathsf{Apx}}$ on the expectation of the total variation distance between $P_s$ and $P_c$. 
Observe that using 
$P_c$ instead of $P_s$ changes the probability of error by at most $\epsilon_{\mathsf{Apx}}$. Thus the expected error probability $\e_{\mb}\mathsf{P}[\me]$ of the channel coding is bounded above by $\ep_{\mathsf{Dec}}+\ep_{\mathsf{Apx}}$.
\subsubsection*{Part 3: Eliminating shared randomness}
Using the law of iterated expectation, we have $\e_{\mb}\mathsf{P}[\me]=\e_{\mb,F}\mathsf{P}[\me|F]\le\ep_{\mathsf{Dec}}+\ep_{\mathsf{Apx}}$. Thus there exists a fixed binning and an instance $f^*$ of $F$, such that the encoder $p_s(x|m,f^*)$ and the mismatched decoder $T(\hat{x}|y,f^*)$ results in a pair of encoder-decoder with error probability of at most $\ep_{\mathsf{Dec}}+\ep_{\mathsf{Apx}}$.
\subsubsection*{Step 2: Non-asymptotic analysis} We would apply the one shot OSRB bound to $n$  i.i.d. repetitions of the DMC $q_{Y|X}$. In \cite{OSRB}, we started from an i.i.d. input for the source coding part. Although using an i.i.d. distribution makes evaluation of Theorem \ref{thm:os1} and Theorem \ref{thm:os2} simple, but this does not yield an optimal strategy. This is due to the fact that an i.i.d. input causes a dispersion in addition to the inherent dispersion of the channel. To avoid input dispersion, we choose channel input sequences with the same type. 

Let $\sm=2^{nR}$ and $\sf=2^{n\tR}$. For a given $q_X$ and $n$, we can find a $n$-type $\Phi^{(n)}_X$ such that the infinity norm $\|\Phi^{(n)}_X-q_X\|_{\infty}\le\frac{1}{n}$. To prove \eqref{eq:ch1}, assume that the $p_{X^n}$ is a uniform distribution over the set $\mt_{\Phi^{(n)}_X}$ of sequences with the type $\Phi^{(n)}_X$. The known bounds on the size of typical sets imply that there exists $L$ such that for sufficiently large $n$, $\log|\mt_{\Phi^{(n)}_X}|\ge nH_{\Phi^{(n)}_X}(X)-L\log n$. Setting $\gamma=\log n$, $Z$ a constant and $|\mathcal{V}|=1$ in Theorem \ref{thm:os1} gives the following bound on the right hand side of equation \eqref{eq:A1} and thus on $\ep_{\mathsf{Apx}}$:
\begin{align}
\ep_{\mathsf{Apx}}\le p_{X^n}(\ms_{\gamma}^c)+\dfrac{1}{\sqrt{n}},\label{eq:13}
\end{align}
where we have used the theorem with $X^n$ being the $X_{\mathcal{V}}$ in the statement of the theorem. Further
\begin{align}
\ms_{\gamma}:=\{x^n:h_{p_{X^n}}(x^n)-\log n>n(R+\tilde{R})\}.
\end{align}
Note that for each $x^n\in\mt_{\Phi^{(n)}_X}$ the relation $h_{p_{X^n}}(x^n)=\log|\mt_{\Phi^{(n)}_X}|$ holds. Hence if we set 
\be n(R+\tilde{R})=nH_{\Phi^{(n)}_X}(X)-(L+2)\log n,\label{eq:15}\ee 
then the first term of \eqref{eq:13} vanishes and we have $\ep_{\mathsf{Apx}}\le \dfrac{1}{\sqrt{n}}$.

Next we should find $\tR$ such that the error probability $\ep_{\mathsf{Dec}}\le\ep-\frac{1}{\sqrt{n}}$. The decoder has access to $Y^n$ and a single bin index $F$ of $X^n$. Setting $\gamma=\dfrac{1}{2}\log n$, $|\mathcal{V}|=1$, $Z=Y^n$ as well as $F$ as a bin index of $X_{\mv}=X^n$ in the statement of Theorem \ref{thm:os2}, we get that for any $t_{X^nY^n}$, we have
\begin{align}
\ep_{\dec}\le p_{X^n}q_{Y^n|X^n}(\ms({t})^c)+\dfrac{1}{\sqrt{n}},\label{eq:14}
\end{align}
where
$
\ms(t):=\{(x^n,y^n):n\tR-h_{{t}}(x^n|y^n)>\dfrac{1}{2}\log n\}
$.
Observe that $Y_1,\cdots,Y_n$ are conditionally independent given any $X^n=x^n$ because the channel is memoryless. So if we can write $h_{t}(x^n|y^n)$ as a sum of independent rv's, we would be able to use Berry-Essen CLT to find $\tR$. Using $t_{X^nY^n}=p_{X^n}q_{Y^n|X^n}$ does not give rise to such a factorization. To overcome this situation we use $t_{X^n,Y^n}=q_{X^n}q_{Y^n|X^n}=\prod_{i=1}^n q(x_i)q(y_i|x_i)$ for mismatch decoding. We then have $h_{t}(x^n|y^n)=\sum_{i=1}^n h_{q}(x_i|y_i)$. Given $X^n=x^n$, $\{h_{q}(x_i|Y_i)\}_{i=1}^n$ are functions of independent rv's, and hence mutually independent; thus we can now apply the Berry-Essen CLT to bound the first term of \eqref{eq:14}. Using the Berry-Essen CLT for each $x^n\in\mt_{\Phi^{(n)}_X}$, we have  
\begin{align}
q_{Y^n|X^n=x^n}(\ms(t)^c)&=p_G(G\ge n\tR-\dfrac{1}{2}\log n)+O(\dfrac{1}{\sqrt{n}})\nonumber
\end{align}  
where $G$ is a normal r.v. with
\begin{align}
&\e G=\sum_{i=1}^n \e_{q_{Y_i|x_i}} h_q(x_i|Y_i)=\sum_{x}\#[x_i=x] \e_{q_{Y|x}} h_q(x|Y)\n
&=
\sum_{x}n\Phi^{(n)}(x) \e_{q_{Y|x}} h_q(x|Y)=n\e_{\Phi^{(n)}_{X}}\e_{q_{Y|X}}[h_q(X|Y)|X]\n
&\Var G=\sum_{i=1}^n \Var_{q_{Y_i|x_i}}h_q(x_i|Y_i)
=n\e_{\Phi^{(n)}_{X}}\Var_{q_{Y|X}}[h_q(X|Y)|X]\n
&~~~~~~~=n\e_{\Phi^{(n)}_{X}}\Var_{q_{Y|X}}[\im_q(X;Y)|X].\nonumber
\end{align} 
The sketch of the rest of the proof is as follows (see \cite{FullVersion} for details): analyzing the bound \eqref{eq:14}, we get that 
\begin{align}
n\tR=n\e_{\Phi^{(n)}_{X}}\e_{q_{Y|X}}[h_q(X|Y)|X]+\sqrt{n\e_{\Phi^{(n)}_{X}}\Var_{q_{Y|X}}[\im_q(X;Y)|X]}Q^{-1}(\ep)+O(\log n),\label{eq:16}\end{align}
is sufficient to achieve $\epsilon_{\mathsf{Dec}}\le\ep-\frac{1}{\sqrt{n}}$. Now $\|\Phi^{(n)}_X-q_X\|_{\infty}\le\frac{1}{n}$ implies that $H_{\Phi^{(n)}_{X}}(X)=H_q(X)+O(\frac{1}{n})$, $\e_{\Phi^{(n)}_{X}}\e_{q_{Y|X}}[h_q(X|Y)|X]=H_q(X|Y)+O(\frac{1}{n})$  and $\e_{\Phi^{(n)}_{X}}\Var_{q_{Y|X}}[\im_q(X;Y)|X]=V_q+O(\frac{1}{n})$ yields 
Finally combining these relations with \eqref{eq:15} and \eqref{eq:16}  imply \eqref{eq:ch1}.
\subsection{Broadcast channel}\label{sec:32}
Consider the problem of transmission of two private messages over a broadcast channel $q_{Y_1Y_2|X}$. Let $\mr^*(n,\ep)$ be the set of all rate pairs $(R_1, R_2)$ of all $(n,\ep)$-codes where $\ep$ is the probability of erroneous decoding at either of the decoders. We prove a one-shot version of Marton with two auxilaries. A similar theorem is proved for Marton with common message and involving auxiliary rv $U_0$ in \cite{FullVersion}. 
\begin{theorem}\label{thm:bc-d}
Given any pmf $q_{U_1U_2X}$, let $\mr_{\mathsf{in}}(q_{U_1U_2X},n,\ep)$ be the set of all pairs $(R_1,R_2)$ for which there exists reals $\tR_1,\tR_2\ge0$ such that
 \begin{align}
 R_j+\tR_j &\le H_q(U_j)-O(\dfrac{\log n}{n}),\ j=1,2,\n
   R_1+R_2+\tR_1+\tR_2&\le H_q(U_1U_2)-O(\dfrac{\log n}{n})\label{eq:30}\\
   \left[\begin{array}{l} \tR_1\\\tR_2\end{array}\right]\in\left[\begin{array}{l} H_q(U_1|Y_1)\\ H_q(U_2|Y_2)\end{array}\right]&+\mq^{-1}(\mathbb{V}_{\mathsf{BC},q},\ep)+O(\dfrac{\log n}{n}),\label{eq:40}
\end{align}
where the entropies are computed according to the pmf $q_{U_1U_2XY_1Y_2}=q_{U_1U_2X}q_{Y_1Y_2|X}$ and 
\[
\mathbb{V}_{\mathsf{BC},q}=\e_{q_{U_1U_2}}\Cov_{q_{Y_1Y_2|U_1U_2}}\left[(\im_q(U_1;Y_1),\im_q(U_2;Y_2))^{\mathsf{T}}|U_1U_2\right].
\]
Then $\cup_{q_{U_1U_2X}}\mr_{\mathsf{in}}(q_{U_1U_2X},n,\ep)\subseteq\mr^*(n,\ep)$.
\end{theorem}
\begin{IEEEproof}[Sketch of the proof]
The proof follows in similar steps as in the proof of channel coding.
\subsubsection*{Part 1: Source coding side of the problem and random binning} Let $(U_1,U_2,X,Y)$ be distributed according to $q(u_1,u_2, x, y_1,y_2)=q(u_1,u_2,x)q(y_1,y_2|x)$. Consider the following random binning
\begin{itemize}
\item For $j=1,2$, to each $u_j$ assign independently two random bins $m_j\in[1:\sm_j]$ and $f_j\in[1:\sf_j]$.
\end{itemize}
 Suppose that the decoder at the receiver $j=1,2$ uses a mismatched decoder $T_j(\hat{u}_j|y_j,f_j)$ to generate $\hat{u}_j$ and thereby $\hat{m}_j$. The induced random pmf is 
 \begin{align} P_s(u_{1:2}&,m_{1:2},f_{1:2},y_{1:2},\hat{u}_{1:2})=q(u_{1:2},y_{1:2})P_s(m_1,f_1|u_1)\n&P_s(m_2,f_2|u_2)T_1(\hat{u}_1|f_1,y_1)T_2(\hat{u}_2|f_2,y_2).\end{align}
We find an upper bound on $\e\mathsf{P}(\hat{u}_1\neq u_1 \mathsf{or}~ \hat{u}_2\neq u_2)$ which in turn bounds the probability of error. Using the first bound of Theorem \ref{thm:os2} and the union bound, we have (see \cite{FullVersion} for proof):
  \begin{lemma}\label{lemma:l1}
For any mismatched decoders $T_j, j=1,2$,
\begin{align}
\e \mathsf{P}[\me]\le& p_{U_{1:2}Y_{1:2}}(\ms_{\gamma}(t_1,t_2)^c)+4\times2^{-\gamma},
\end{align}
where $\gamma$ is an arbitrary positive number and 
\begin{align}
\ms_{\gamma}(t_1,t_2):=\{(u_{1:2},y_{1:2}):\log \sf_j-h_{t_j}(u_j|y_j)>\gamma, j=1,2\}.
\end{align}

  \end{lemma} 
\subsubsection*{Part 2: Designing encoder-decoder assisted with a shared randomness}
Assume that there is a shared randomness $(F_1,F_2)$ available at the both encoders and the decoder, which is independent of the message and uniformly distributed over $[1:\sf_1]\times[1:\sf_2]$. The encoder uses the conditional pmf $P_s(u_{1:2},x|m_{1:2},f_{1:2})$ of the source coding problem. The decoder $j$ uses the mismatched decoder $T_j(\hat{u}_j|y_j,f_j)$ to find $\hat{u}_j$ and as a result an estimate $\hat{m}_j$ of the message. Then the induced random pmf is $P_c(u_{1:2},x,y_{1:2},m_{1:2},f_{1:2},\hat{u}_{1:2})=p^U(m_{1:2},f_{1:2})P_s(u_{1:2},x,y_{1:2},\hat{u}_{1:2}|m_{1:2},f_{1:2})$. We have 
\[\tv{P_s-P_c}=\tv{P_s(m_{1:2},f_{1:2})-p^{U}(m_{1:2},f_{1:2})}.\]
The probability of error is no more than $\tv{P_s-P_c}$ and thus no more than the right hand side of the above equation.
Given $\sm$ and $\sf$, Theorem \ref{thm:os1} gives an upper bound $\epsilon_{\mathsf{Apx}}$ on the expectation of the right hand side.
Observe that the expected error probability $\e_{\mb}\mathsf{P}[\me]$ of the channel coding is bounded from above by $\ep_{\mathsf{Dec}}+\ep_{\mathsf{Apx}}$. 

Finally we can eliminate the shared randomness $F_{1:2}$ as in the proof of the channel coding.
\subsubsection*{Step 2: Non-asymptotic analysis of one-shot OSRB} We would apply the one shot OSRB bound to $n$ repetitions of the BC $q_{Y_{1:2}|X}$.

Let $\sm_j=2^{nR_j}$ and $\sf_j=2^{n\tR_j}$. Following the proof of channel coding, for a given $q_X$ and $n$, we find an $n$-type $\Phi^{(n)}_{U_{1:2}}$ such that $\|\Phi^{(n)}_{U_{1:2}}-q_{U_{1:2}}\|_{\infty}\le\frac{1}{n}$. To prove \eqref{eq:ch1}, assume that $p_{U_{1:2}^n}$ is a uniform distribution over the set $\mt_{\Phi^{(n)}_{U_{1:2}}}$ of sequences with the type $\Phi^{(n)}_{U_{1:2}}$. Observe that $p_{U_j^n}$ has a uniform distribution over the set $\mt_{\Phi^{(n)}_{U_{j}}}$ of sequences with the type $\Phi^{(n)}_{U_{j}}$. As a result, $h_{p_{U_{1:2}^n}}(u_j^n)=\log|\mt_{\Phi^{(n)}_{U_{j}}}|$. As in the proof of channel coding, we can utilize Theorem \ref{thm:os1}  to show that if  the inequalities in \eqref{eq:30} are satisfied, then  $\ep_{\mathsf{Apx}}\le O(\dfrac{1}{\sqrt{n}})$.

Next we should find $\tR_1,\tR_2$ such that $\ep_{\dec}\le\ep-O(\frac{1}{\sqrt{n}})$. We can utilize Lemma \ref{lemma:l1} to show that \eqref{eq:40} is sufficient for $\ep_{\dec}\le\ep-O(\frac{1}{\sqrt{n}})$. The rest of the proof is similar to that of channel coding but uses a
generalized version of Berry-Essen CLT for the independent and multidimensional r.v.'s \cite{LyCLT}. \end{IEEEproof}
\subsection{Wiretap channel with strong secrecy}\label{sec:33}
Consider a wiretap channel with probability transition $q_{YZ|X}$, in which the receiver and the wiretapper have access to channel outputs $Y$ and $Z$, respectively. For a given $(n,R)$ code we use total variation distance $\tv{p_{MZ^n}-p^{U}_Mp_{Z^n}}$ to measure the security of the code, where $p_{MZ^n}$ is the induced pmf by the code. A rate $R$ is said to be $(\ep_r,\ep_{\sec})$-achievable if there exists an $(n,R)$ code such that $\mathsf{P}[\me]\le\ep_r$ and $\tv{p_{MZ^n}-p^{U}_Mp_{Z^n}}\le\ep_{\sec}$.
\begin{theorem}
Given $q_{Y,Z|X}$, for any input distribution $q_{U,X}$ and any $\theta\in[0,1]$, the following rate is $(n,\ep_r,\ep_{\sec})$-achievable:
\begin{align}\label{eq:74}
R(n,\ep_r,\ep_{\sec})=I_q(U;Y)-I_q(U;Z)-\sqrt{n\mathsf{V}_Y}Q^{-1}(\theta\ep_r)-\sqrt{n\mathsf{V}_Z}Q^{-1}(\bar{\theta}\ep_{\sec})-O(\dfrac{\log n}{n}),
\end{align}
where $\bar{\theta}=1-\theta$, $\mathsf{V}_Y=\e_{q_{UX}}\Var_{q_{Y|UX}}[I_q(U;Y)|U]$ and $\mathsf{V}_Z$ is defined similarly.
\end{theorem}
\begin{IEEEproof}[Sketch of proof]
For simplicity, we prove the theorem for the special case $U=X$. We will find $R(n,\ep_r,\ep_{\sec})$ such that $\e\mathsf{P}[\me]\leq \theta\ep_r$ and $\e\tv{p_{MZ^n}-p^{U}_Mp_{Z^n}}<\bar{\theta}\ep_{\sec}$. Then by Markov inequality, we can find a code with the desired conditions.  
\subsubsection*{One-shot  OSRB} We use the same code construction of subsection \ref{sub:ch}. Here we need to compute the security index of the code. To do this, we bound the security constraint, i.e. $\ep_{s,\sec}=\e\tv{P_s(m,f,z)-p^U(m,f)q(z)}$, using Theorem \ref{thm:os1} in the source coding part of the problem. Then using triangular inequality, the security constraint of channel coding asserted with shared randomness $\ep_{c,\sec}=\e\tv{P_c(m,f,z)-p^U(m,f)q(z)}$ is bounded above by $\ep_{s,\sec}+\ep_{\ap}$. To eliminate $F$, we can show that there exists an instance $f$ such that $\e_{\mb|F=f}\tv{P_c(m,z|f)-p^U(m)P_c(z)}\le \ep_{s,\sec}+3\ep_{\ap}$. 
\subsubsection*{Non-asymptotic analysis of one-shot OSRB} Again we follow the analysis of pt-to-pt channel. It was observed that if \eqref{eq:15} is satisfied, then $\ep_{\ap}\le 1/\sqrt{n}$. Similar error analysis shows that $\e\mathsf{P}[\me]\le\theta\ep_r$ provided that
\begin{align}
n\tR=nH_q(X|Y)+\sqrt{n\mathsf{V}_Y}Q^{-1}(\theta\ep_r)+O(\log n).\label{eq:last-1}
\end{align} 
Next we find a constraint on $R$ and $\tR$ such that  security index $\ep_{s,\sec}\le \bar{\theta}\ep_{\sec}-3/{\sqrt{n}}$, which shows that $\e\tv{p_{MZ^n}-p^{U}_Mp_{Z^n}}<\bar{\theta}\ep_{\sec}$. Substituting $\gamma=\log n$ in Theorem \ref{thm:os1} gives:
$
\ep_{s,\sec}\le {p}_{X^n}(\ms_{\gamma}^c(p\|t))+1/{\sqrt{n}},
$
where
\begin{align}
\ms_{\gamma}(p\|t):=\{(x^n,z^n):h_{p_{X^n}}(x^n,z^n)-h_t(z^n)-\log n>n(R+\tilde{R})\}.\nonumber
\end{align}
Again as in the proof of error probability for channel coding, to apply Berry-Essen CLT we need to write $h_{p_{X^n}}(x^n,z^n)-h_t(z^n)$ as a sum of independent r.v.'s.  Let $t(z^n)=\prod_{i=1}^n q_Z(z_i)$, which makes $h_t(z^n)$ as sum of independent r.v.'s. Next observe that for any $x^n\in\mt_{\Phi_{X^n}}$, $p_{X^n}q_{Z^n|X^n}(x^n,z^n)=q_{X^n,Z^n}(x^n,z^n)2^{-O(\log n)}$. Using this fact, we have $h_{p_{X^n}q_{Z^n|X^n}}(x^n,z^n)=h_q(x^n,z^n)+O(\log n)$; thus
\begin{align*}
\ms_{\gamma}(p\|t):=\{(x^n,z^n):\sum_{i=1}^n h_q(x_i|z_i)-O(\log n)>n(R+\tilde{R})\}.
\end{align*}
Applying Berry-Essen CLT in the same way as in channel coding proof implies
\begin{align}
n(R+\tR)=nH_q(X|Z)-\sqrt{n\mathsf{V}_Z}Q^{-1}(\bar{\theta}\ep_{\sec})+O(\log n).\label{eq:last}
\end{align} 
Combining \eqref{eq:last-1} and \eqref{eq:last} yields \eqref{eq:74}.
\end{IEEEproof}

\appendices
\section{Proof of Theorem \ref{thm:os1}}\label{apx:nosrb}
We modify the proof of Theorem 4 of \cite{OSRB} to obtain the one-shot version of OSRB. Observe that 
\bes
\label{eq:apx}
\e P(z,b_{\mv})=\frac{1}{\prod_{v\in\mv}\sm_v}p(z)=p^U(b_{\mv})p(z).
\ees

We can decompose this sum into two parts  
 as follows:
\begin{align*}
\hat{P}(z,b_{\mv})&=\sum_{x_{\mv}: (x_{\mv},z)\in \ms_{\gamma}(p\|t)}p(x_{\mv},z)\prod_{v\in\mv}\mathbf{1}\{\mb_v(x_v)=b_v\},\\
\tilde{P}(z,b_{\mv})&=\sum_{x_{\mv}: (x_{\mv},z)\notin \ms_{\gamma}(p\|t)}p(x_{\mv},z)\prod_{v\in\mv}\mathbf{1}\{\mb_v(x_v)=b_v\}.
\end{align*}
We then have
$$P(z,b_{\mv})=\hat{P}(z,b_{\mv})+\tilde{P}(z,b_{\mv}).$$
Thus
$$\e P(z,b_{\mv})=
\e \hat{P}(z,b_{\mv})+
\e \tilde{P}(z,b_{\mv}).$$

Using triangle inequality we have
\begin{align}
\e\tv{P(z,b_{\mv})-p_{Z}(z)\prod_{v\in\mv} p^U(b_t)}&=\e\tv{P(z,b_{\mv})-\e P(z,b_{\mv})}\n&\leq
\e\tv{\hat{P}(z,b_{\mv})-\e \hat{P}(z,b_{\mv})}\n&\qquad+
\e\tv{\tilde{P}(z,b_{\mv})-\e \tilde{P}(z,b_{\mv})}.\label{eq:A1}
\end{align}
Therefore we need to show that both of the terms on the right hand side converge to zero as $n$ converges to infinity.
For the second term we have
\begin{align}
\e\tv{\tilde{P}(z,b_{\mv})-\e \tilde{P}(z,b_{\mv})}&\leq \frac{1}{2}\sum_{z,b_{\mv}}2\e \tilde{P}(z,b_{\mv})\n
                                                                                    &= \e\sum_{z,x_{\mv},b_{\mv}:(z,x_{\mv})\notin\styp} P(z,x_{\mv},b_{\mv})\n
                                                                                    &=\e\sum_{z,x_{\mv}:(z,x_{\mv})\notin\styp} P(z,x_{\mv})\n
                                                                                    &=\e\sum_{z,x_{\mv}:(z,x_{\mv})\notin\styp} p(z,x_{\mv})\n
                                                                                    &=p_{X_{\mv}Z}\left((\styp)^c\right),
\end{align}
where the factor $\frac 12$ in the first equation comes from the definition of total variation distance. In the first step we use the inequality $\e|X-\e X|\le 2\e|X|$ and in the last step, we use the relation $P(z,x_{\mv})=p_{X_{\mv}Z}(z,x_{\mv})$ since $P(z,x_{\mv})$ is not a random pmf.

Next consider the first term of the r.h.s.\ of equation \eqref{eq:A1}. Using Cauchy-Schwarz inequality we have  \be\label{eq:A4.5}
\e\tv{\hat{P}(z,b_{\mv})-\e \hat{P}(z,b_{\mv})}=\frac{1}{2} \sum_{z,b_{\mv}}\mathbb{E}|\hat{P}(z,b_{\mv})-\e\hat{P}(z,b_{\mv})| \le\frac{1}{2}\sum_{z,b_{\mv}}\sqrt{\mathsf{var}(\hat{P}(z,b_{\mv}))}.
\ee
Using the formula $\mathsf{var}(\sum_{i=1}^TX_i)=\sum_{1\le i,j\le T}\mathsf{cov}(X_i,X_j)$ and the fact that $\hat{P}(z,b_{\mv})$ is sum of several terms  
\begin{align}
\hat{P}(z,b_{\mv})&=\sum_{x_{\mv}: (x_{\mv},z)\in \ms_{\gamma}(p\|t)}p(x_{\mv},z)\prod_{v\in\mv}\mathbf{1}\{\mb_v(x_v)=b_v\},
\end{align}
we can write
\begin{align}
\mathsf{var}(\hat{P}(z,b_{\mv}) )    &=\sum_{x_{\mv},\bar{x}_{\mv}:\atop {(x_{\mv},z)\in\styp,\  (\bar{x}_{\mv},z)\in\styp}}\mathsf{cov}\Big(p(x_{\mv},z)\mathbf{1}\{\mb_{\mv}(x_{\mv})=b_{\mv}\},p(\bar{x}_{\mv},z)\mathbf{1}\{\mb_{\mv}(\bar{x}_{\mv})=b_{\mv}\}\Big)\n
                                                         &=\sum_{x_{\mv},\bar{x}_{\mv}:\atop {(x_{\mv},z)\in\styp,\  (\bar{x}_{\mv},z)\in\styp}}p(x_{\mv},z)p(\bar{x}_{\mv},z)\mathsf{cov}\Big(\mathbf{1}\{\mb_{\mv}(x_{\mv})=b_{\mv}\},\mathbf{1}\{\mb_{\mv}(\bar{x}_{\mv})=b_{\mv}\}\Big)\label{eq:A46}
\end{align}
To evaluate the \emph{covariance} term, we need to find the places where $x_{\mv}$ and $\bar{x}_{\mv}$ match.  We first partition the set $\{(x_{\mv},\bar{x}_{\mv}): (x_{\mv},z)\in\styp,\  (\bar{x}_{\mv},z)\in\styp\}$ into the sets $\mn(\mbox{\footnotesize$\ms$},z)$ defined below:
\begin{align*}
\mn(\mbox{\footnotesize$\ms$},z):=\Big\{(x_{\mv},\bar{x}_{\mv}): &x_{\ms}=\bar{x}_{\ms},\ x_{t}\neq\bar{x}_{t}, \forall t\in\mbox{\footnotesize$\ms$}^c,\\&
\ (x_{\mv},z)\in\styp,\  (\bar{x}_{\mv},z)\in\styp\Big \}.
\end{align*}
We note that for each pair $(x_{\mv},\bar{x}_{\mv})$ inside the set $\mn(\emptyset,z)$, the random variables $\mathbf{1}\{\mb_{\mv}^{(n)}(x_{\mv})=b_{\mv}\}$ and $\mathbf{1}\{\mb_{\mv}^{(n)}(\bar{x}_{\mv})=b_{\mv}\}$ are independent, thus the covariance between them is \underline{zero}. Next for $\mbox{\footnotesize$\ms$}\neq\emptyset$, we bound above the covariance for the pair $(x_{\mv},\bar{x}_{\mv})\in \mn(\mbox{\footnotesize$\ms$},z)$ as follows,
\begin{align}
\mathsf{cov}\left(\mathbf{1}\{\mb_{\mv}^{(n)}(x_{\mv})=b_{\mv}\},\mathbf{1}\{\mb_{\mv}^{(n)}(\bar{x}_{\mv})=b_{\mv}\}\right)
                                                          &\le\e\ \mathbf{1}\{\mb_{\mv}^{(n)}(x_{\mv})=\mb_{\mv}^{(n)}(\bar{x}_{\mv})=b_{\mv}\}\n
                                                          &=\e\ \mathbf{1}\{\mb_{\ms}^{(n)}(x_{\ms})=b_{\ms},\mb_{\ms^c}^{(n)}(x_{\ms^c})=\mb_{\ms^c}^{(n)}(\bar{x}_{\ms^c})=b_{\ms^c}\}\n
                                                          &=\frac{1}{\sm_{\ms}\sm_{\ms^c}^2},\label{eq:A47}
\end{align}
where in the last step, we use the fact that the random variables $\mathbf{1}\{\mb_{\ms}^{(n)}(x_{\ms})=b_{\ms}\}$,  $\mathbf{1}\{\mb_{\ms^c}^{(n)}(x_{\ms^c})=b_{\ms^c}\}$ and $\mathbf{1}\{\mb_{\ms^c}^{(n)}(\bar{x}_{\ms^c})=b_{\ms^c}\}$ are mutually independent.

Since the union is taken over $x_{\mv}$ and $\bar{x}_{\mv}$ in the set $\styp$, we need to get access to the set $\mbox{\footnotesize$\ms$}$ that appears in the definition of $\styp$. For this reason for any subset $\mbox{\footnotesize$\ms$}$ we define
\begin{align}
\styp(\mbox{\footnotesize$\ms$}):=\left\{(x_{\ms},z): h_p(x_{\ms},z)-h_t(z)-\sum_{v\in\ms}\log \sm_v>\gamma, \right\}.
\end{align}
Then from the definition of $\styp$ we have $\styp=\bigcap_{\ms\subseteq\mv}\{x_{\mv}:x_{\ms}\in \styp(\mbox{\footnotesize$\ms$})\}$. 
Substituting \eqref{eq:A47} in \eqref{eq:A46} gives
\begin{align}
\mathsf{var}(\hat{P}(z,b_{\mv}))&\le \sum_{\emptyset\neq\ms\subseteq\mv}\sum_{(x_{\mv},\bar{x}_{\mv})\in\mn(\ms,z)}p(x_{\mv},z)p(\bar{x}_{\mv},z)\frac{1}{\sm_{\ms}\sm_{\ms^c}^2}\n
&\le \sum_{\emptyset\neq\ms\subseteq\mv}\frac{1}{\sm_{\ms}\sm_{\ms^c}^2}\sum_{x_{\ms}:(x_{\ms},z)\in\styp(\ms)}p^2(x_{\ms},z)\n
&~~~~~~~~~~~~~~~~~~\times \sum_{x_{\ms^c},\bar{x}_{\ms^c}}p(x_{\ms^c}|x_{\ms},z)p(\bar{x}_{\ms^c}|x_{\ms},z)\label{eqn:TOZIH}\\
&= \sum_{\emptyset\neq\ms\subseteq\mv}\frac{1}{\sm_{\ms}\sm_{\ms^c}^2}\sum_{x_{\ms}:(x_{\ms},z)\in\styp(\ms)}p^2(x_{\ms},z) \n
&\le \sum_{\emptyset\neq\ms\subseteq\mv}\frac{1}{\sm_{\ms}\sm_{\ms^c}^2}\sum_{x_{\ms}:(x_{\ms},z)\in\styp(\ms)}p(x_{\ms},z)t(z)\frac{1}{\sm_{\ms}}2^{-\gamma} \label{eq:AAA}\\
&\le \sum_{\emptyset\neq\ms\subseteq\mv}\frac{1}{\sm_{\ms}\sm_{\ms^c}^2}\sum_{x_{\ms}}p(x_{\ms},z)t(z)\frac{1}{\sm_{\ms}}2^{-\gamma} \label{eq:AAA1h}\\
&\le \frac{1}{\sm_{\mv}^2}2^{\mv-\gamma}p(z)t(z)\label{eq:AAA1}
\end{align}
where \eqref{eqn:TOZIH} follows from relaxing the restrictions $x_{\mv}\in\styp$ and $\bar{x}_{\mv}\in \styp$ while keeping the constraint $\bar{x}_{\ms}=x_{\ms}\in\styp(\mbox{\footnotesize$\ms$})$. Eq. \eqref{eq:AAA} follows from $p(x_{\ms},z)\leq t(z)\frac{1}{\sm_{\ms}}2^{-\gamma}$ which is due to 
the definition of the set $\styp(\mbox{\footnotesize$\ms$})$.

Substituting \eqref{eq:AAA1} in \eqref{eq:A4.5} yield
\begin{align}
\frac{1}{2}\sum_{z,b_{\mv}}\sqrt{\mathsf{var}(\hat{P}(z,b_{\mv}))}&\le \sum_{z,b_{\mv}} \frac{1}{\sm_{\mv}}2^{\frac{\mv-\gamma}{2}-1}\sqrt{p(z)t(z)}\n
&=2^{\frac{\mv-\gamma}{2}-1} \sum_{z} \sqrt{p(z)t(z)}\n
&\le 2^{\dfrac{\mv-\gamma}{2}-1},\label{eq:AAAA}
\end{align}
where \eqref{eq:AAAA} follows from Cauchy-Schwarz inequality. 
\section{Proof of Theorem \ref{thm:os2}}\label{apx:sw}
First we prove the inequality \eqref{eq:sw} following the proof for the special case $|\mv|=1$. We have
\[\mathsf{P}[C]=\sum_{x_{\mv},b_{\mv},z}p(x_{\mv},z)\mathbf{1}(B_{\mv}(x_{\mv})=b_{\mv})Q_{X_{\mv}|Z,B_{\mv}}(x_{\mv}|z,b_{\mv}).\]
Consider,

\begin{IEEEeqnarray}{rCl}
\e\mathsf{P}[C]&=&\e\sum_{x_{\mv},z,b_{\mv}}p(x_{\mv},z)\mathbf{1}(B(x_{\mv})=b_{\mv})\dfrac{t(x_{\mv}|z)}{\sum_{\bar{x}_{\mv}}t(\bar{x}_{\mv}|z)\mathbf{1}(B(\bar{x}_{\mv})=b_{\mv})}\\
                         &=&\sm_{\mv}\e\sum_{x_{\mv},z}p(x_{\mv},z)\mathbf{1}(B(x_{\mv})=1_{\mv})\dfrac{t(x_{\mv}|z)}{\sum_{\bar{x}_{\mv}}t(\bar{x}_{\mv}|z)\mathbf{1}(B(\bar{x}_{\mv})=1_{\mv})}\label{eq:SW01}\\
                         &=&\sm_{\mv}\sum_{x_{\mv},z}\e_{B(x_{\mv})}\e_{\{B(\bar{x}_{\mv}), \bar{x}_{\mv}\neq x_{\mv}\}}p(x_{\mv},z)\dfrac{t(x_{\mv}|z)\mathbf{1}(B(x_{\mv})=1_{\mv})}{\sum_{\bar{x}_{\mv}}t(\bar{x}_{\mv}|z)\mathbf{1}(B(\bar{x}_{\mv})=1_{\mv})}\label{eq:SW02}\\
                         &\ge&\sm_{\mv}\sum_{x_{\mv},z}\e_{B({x}_{\mv})}p(x_{\mv},z)\dfrac{t(x_{\mv}|z)\mathbf{1}(B(x_{\mv})=1_{\mv})}{\e_{\{B(\bar{x}_{\mv}), \bar{x}_{\mv}\neq x_{\mv}\}}\iffalse\e_{\mathcal{B}|\mathbf{1}(B(x_{\mv})=1_{\mv})}\fi\sum_{\bar{x}_{\mv}}t(\bar{x}_{\mv}|z)\mathbf{1}(B(\bar{x}_{\mv})=1_{\mv})}\label{eq:SW03}\\
                         &\ge&\sm_{\mv}\sum_{x_{\mv},z}\e_{B(x_{\mv})}p(x_{\mv},z)\dfrac{t(x_{\mv}|z)\mathbf{1}(B(x_{\mv})=1_{\mv})}{\sum_{\ms\subseteq\mv}\sm_{\ms}^{-1}t({x}_{\ms^c}|z)\mathbf{1}(B({x}_{\ms^c})=1_{\ms})}\label{eq:SW04}\\                                                  
                         &=&\sum_{x_{\mv},z}p(x_{\mv},z)\dfrac{t(x_{\mv}|z)}{\sum_{\ms\subseteq\mv}\sm_{\ms}^{-1}t({x}_{\ms^c}|z)}\label{eq:SW05}\\
                         &=&\e_{p}\dfrac{1}{1+\sum_{\emptyset\neq\ms\subseteq\mv}\sm_{\ms}^{-1}2^{h_t(X_{\ms}|X_{\ms^c},Z)}},  \label{eq:SW06}                
\end{IEEEeqnarray}
where \eqref{eq:SW01} is due to the symmetry and \eqref{eq:SW03} follows from the Jensen inequality for the convex function $f(x)=\frac{1}{x}$ on the $\mathbb{R}_+$. To obtain \eqref{eq:SW04}, we partition the tuples in the set $\mx_{\mv}$ according to its difference with the tuple $x_{\mv}$. Define $\mn_{\ms}:=\{\bar{x}_{\mv}:\bar{x}_{\ms^c}=x_{\ms^c}, \forall v\in\ms:\ \bar{x}_v\neq x_v\}$. Then $\mx_{\mv}=\cup_{\ms\subseteq\mv}\mn_{\ms}$ and for each $\bar{x}_{\mv}\in\mn_{\ms}$, we have
\bes
\e_{\{B(\bar{x}_{\mv}), \bar{x}_{\mv}\neq x_{\mv}\}}\mathbf{1}(B(\bar{x}_{\mv})=1_{\mv})=\sm_{\ms}^{-1}\mathbf{1}(B({x}_{\ms^c})=1_{\ms^c}),
\ees
where we have used the fact that $[B(\bar{x}_v):v\in\ms]$ and $B(x_{\mv})$ are mutually independent. This implies \eqref{eq:SW04}.

\subsubsection*{Weakening the bound \eqref{eq:sw}} 

\begin{IEEEeqnarray}{rCl}
\e_{p}\dfrac{1}{1+\sum_{\emptyset\neq\ms\subseteq\mv}\sm_{\ms}^{-1}2^{h_t(X_{\ms}|X_{\ms^c},Z)}}&\ge&\e_{p}\dfrac{\mathbf{1}\left((X_{\mv},Z)\in\ms_{\gamma}(t_{X_{\mv},Z})\right)}{1+\sum_{\emptyset\neq\ms\subseteq\mv}\sm_{\ms}^{-1}2^{h_t(X_{\ms}|X_{\ms^c},Z)}}\n
&\ge&\e_{p}\dfrac{\mathbf{1}\left((X_{\mv},Z)\in\ms_{\gamma}(t_{X_{\mv},Z})\right)}{1+\sum_{\emptyset\neq\ms\subseteq\mv}2^{-\gamma}}\label{eq:A1A}\\
&=&\e_{p}\dfrac{\mathbf{1}\left((X_{\mv},Z)\in\ms_{\gamma}(t_{X_{\mv},Z})\right)}{1+(2^{\mv}-1)2^{-\gamma}}\n
&=&\dfrac{1}{1+(2^{\mv}-1)2^{-\gamma}}p_{X_{\mv}Z}\left(\ms_{\gamma}(t_{X_{\mv},Z})\right),\label{eq:A2A}
\end{IEEEeqnarray}
where \eqref{eq:A1A} follows from the definition of $\ms_{\gamma}(t_{X_{\mv},Z})$. Finally we have
\newpage
\begin{IEEEeqnarray}{rCl}
\mathsf{P}[\me]&=&1-\mathsf{P}[C]\n
                         &\le&p_{X_{\mv}Z}\left(\ms_{\gamma}(t_{X_{\mv},Z})^c\right)+\left(1-\dfrac{1}{1+(2^{\mv}-1)2^{-\gamma}}\right)p_{X_{\mv}Z}\left(\ms_{\gamma}(t_{X_{\mv},Z})\right)\label{eq:A3A}\\
                         &\le&p_{X_{\mv}Z}\left(\ms_{\gamma}(t_{X_{\mv},Z})^c\right)+\left(1-\dfrac{1}{1+(2^{\mv}-1)2^{-\gamma}}\right)\n
                         &\le&p_{X_{\mv}Z}\left(\ms_{\gamma}(t_{X_{\mv},Z})^c\right)+(2^{\mv}-1)2^{-\gamma},
\end{IEEEeqnarray}
where \eqref{eq:A3A} follows from \eqref{eq:A2A}.

\end{document}